\documentclass[12p]{elsart}
\usepackage{natbib}
\usepackage[pdftex,dvips,xdvi]{graphicx}
\usepackage{subfigure}
\usepackage[figuresleft]{rotating}
\usepackage{amssymb}
\usepackage{thumbpdf}
\usepackage[dvipdf,colorlinks,citecolor=blue]{hyperref}

\hypersetup{%
    pdftitle= {Compass Rose},
    pdfsubject= {Working Paper, Finance},
    pdfkeywords= {Price Clustering and Discreteness, Microstructure,  Compass
    rose, Nonlinear Dynamics},
    pdfauthor= {\textcopyright\ Costas Vorlow},
    pdfcreator= {\LaTeX\ with package \flqq hyperref\frqq},
    pdfproducer= {dvipdf}
    }

\begin{document}

\begin{frontmatter}

\title{Stock Price Clustering and Discreteness: The ``Compass Rose" and Predictability}

\author[Durham]{Constantinos E. Vorlow\corauthref{cor1}}
\ead{Costas@vorlow.org} \ead[url]{http://www.vorlow.org}
\address[Durham]{Durham Business School, University of Durham, Mill Hill Lane, Durham, DH13LB, UK.}
\corauth[cor1]{Corresponding author}

\journal{XXX Preprint Archive}

\begin{abstract}
In this letter we  investigate the information provided by the
``compass rose" (Crack, T.F. and Ledoit, O. (1996), Journal of
Finance, 51(2), pg. 751-762) patterns revealed in phase portraits of
daily stock returns. It has been initially suggested that the
compass rose is just  a manifestation of price clustering and
discreteness and the tick size, factors that can affect the
unbiasedness of an array of statistical tests based on stock
returns. We show that this may not entirely be  the case.
\end{abstract}

\end{frontmatter}


\section{Introduction}

Price clustering and discreteness is considered an important chapter
of the ``market microstructure" literature with serious implications
for tests of market efficiency, risk evaluation techniques and
optimal design of securities procedures. The ``\textit{compass
rose}", introduced dy \citet{Crack96}, is simply the manifestation
of price clustering and discreteness in two or three dimensional
phase portraits. More precisely, a pattern is usually revealed in
scatter diagrams of daily percentage returns against their lagged
values: rays emanating from the center of the portrait (the origin
of the Cartesian axis system), generating a compass rose like
formation of clusters of points.  \citet{Crack96} identify three
conditions necessary for the appearance of the compass rose pattern:
\begin{enumerate}
    \item The daily price changes of the stock should be small
    relative to the price level;
    \item daily price changes should also be effected in discrete
    jumps of a small number of ticks and
    \item the price of the stock should vary over a relatively
    wide range.
\end{enumerate} If any of the above points is violated, compass
rose patters will be very weak or fail to realize altogether.

Several papers have recently appeared on the compass rose theme.
Some mainly confirm the compass rose as a result of the tick size,
price clustering and discreteness and investigate or relax the
above conditions under which it appears (see \citealp{Szpiro98},
\citealp{Wang00}, \citealp{Chen97}, \citealp{Lee99},
\citealp{Gleason00}, \citealp{Wang02} and \citealp{McKenzie:03}) .
Others (see \citealp{Fang02}, \citealp{Kramer97},
\citealp{Koppl:01} and \citealp{Amilon:03}) also concentrate on
how the above factors affect  the validity of statistical tests
based on stock return sequences that reveal this ``nanostructure"
\citep{Szpiro98}. The compass rose should have appeared in several
other works (such as \citealp{EnrightPhD}, \citealp{Chen93},
\citealp{kar95}, \citealp{BM:96}, \citealp{Franses98:book},
\citealp{Franses00:book} and \citealp{kar00}). However this was
not achieved due to an unfortunate choice of graphical
representation style or resolution and of the length (history) of
the sequences examined.

\citet{Crack96} suggested originally that the compass rose may not
offer any help in predicting returns series. This is due to an
apparent  absence of any strong temporal continuity pattern  in
the phase portraits (see Fig. \ref{fig:fig1}a). In this letter we
show that from a small manipulation of the information that
appears in the compass roses, we can gain additional information
on the dynamics of stock returns processes in a very costless
algorithmically way. An initial analysis of this information
suggests the presence of strong nonlinear and possibly
deterministic dynamics.

\section{A different view of the compass rose}

\begin{figure}[ht]
  \centering
    \subfigure[Compass
    Rose.]{\includegraphics*[width=2.6in]{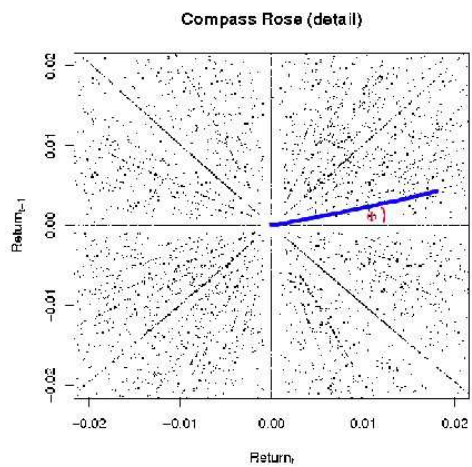}}
    \subfigure[Arc Distribution.]{\includegraphics*[width=2.6in]{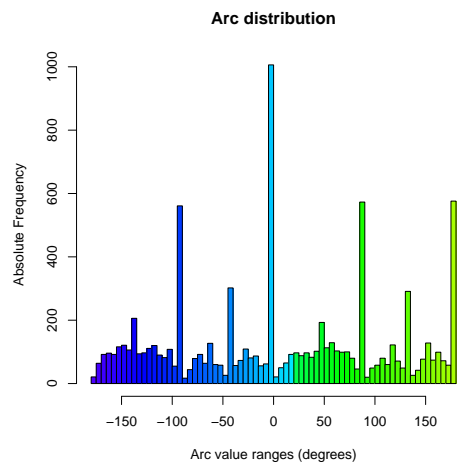}}\\ \quad
   \subfigure[Arc values (unsorted).]{\includegraphics*[width=2.6in]{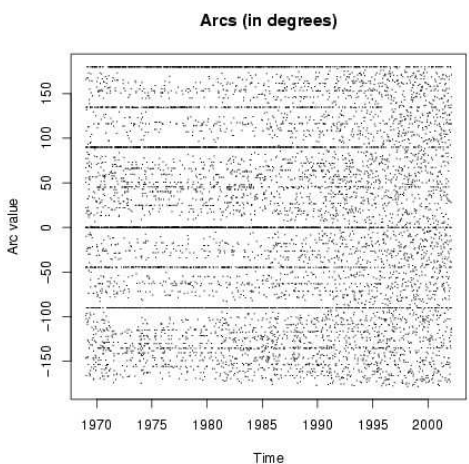}}
   \subfigure[Arc values
   (sorted).]{\includegraphics*[width=2.6in]{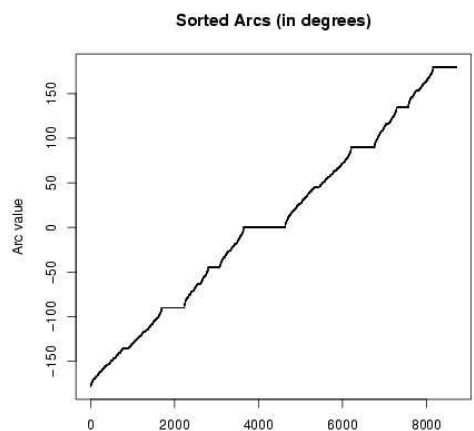}}\\ \quad
  \caption{Patterns observed in the arc distribution.}\label{fig:fig1}
\end{figure}

Price clustering and discreteness, as manifested in the compass
rose, appears in the form of rays that emanate from the center of
the phase portraits (see Fig. \ref{fig:fig1}(a) were we have a
detail of the compass rose for the returns of the TESCO stock,
chosen randomly from stocks forming to the FTSE100 index). It has
been suggested that there is no predictability in such phase
portraits \citep{Crack96}. This is due to temporal structure and
information being concealed in the pase portraits of return
sequences. One could initially adopt the view that the
distribution of the points in the compass rose in Fig.
\ref{fig:fig1} could have been generated by a suitably discretized
random Gaussian process. However \citep{Vorlow:03} showed that
there may be a more delicate temporal pattern hidden under noise
in the compass rose.

\medskip
\begin{center} [ Insert Fig. \ref{fig:fig1} about here. ]
\end{center}
\medskip

In this letter we follow the approach of \citet{Koppl:01} and
instead of examining directly the distribution of points in the
phase portraits, we choose to model the value of the arcs $\Phi$
formed by the line joining the point with the origin (0,0) and the
horizontal axis (Fig. \ref{fig:fig1}a). We measure this in degrees
for simplicity. An advantage of such ``recoding" of the compass
rose information is that all prices will range between 0 and
2$\pi$ or 0 and 360 degrees. It will also allow us to observe more
efficiently the clustering and obtain a non-subjective view of the
compass rose pattern (refer to \citealp{Crack96} and
\citealp{Koppl:01} for discussions).  Indeed, the investigation of
the distribution of the compass rose arc values for TESCO (Fig.
\ref{fig:fig1}b), shows that there is concentration of points
across the horizontal and vertical axis (0, 90, 180 and 270 (-90)
degrees) and  the diagonal rays (close to 45, 45+90, 45+180 and
-45 degrees i.e., the NE-SW and NW-SE directions of the compass
rose). This is referred to as \emph{X-skewness} in
\citet{Koppl:01} and attributed to the presence of big-players and
herding (see \citealp{Koppl:96a} and \citealp{Koppl:96b}). They
also suggested that X-skewing may be inconsistent with ARCH
effects and demonstrate this with simulations.\footnote{The
multimodality of the distribution in the histogram of Fig.
\ref{fig:fig1}(b) may also be an indication of more complex
dynamics. See in \citep{Kantz} and \citep{Kaplan95} for
discussions.} \cite{Mand:99,Mandsciam:99} also revealed a similar
result through a different approach. Judging from the above
literature, it seems that several differently fabricated nonlinear
and non-random sequences can pass as (G)ARCH processes. There is
also the case that stock returns may be characterized by more
complex dynamics, not excluding deterministic or chaotic
structures (refer to \citealp{Kyrtsou2,Kyrtsou1} and
\citealp{Vorlow:00,Vorlow:03,Vorlow:04} for more details).

\begin{figure}
  \centering
    \subfigure[Lag 1.]{\includegraphics*[width=4in]{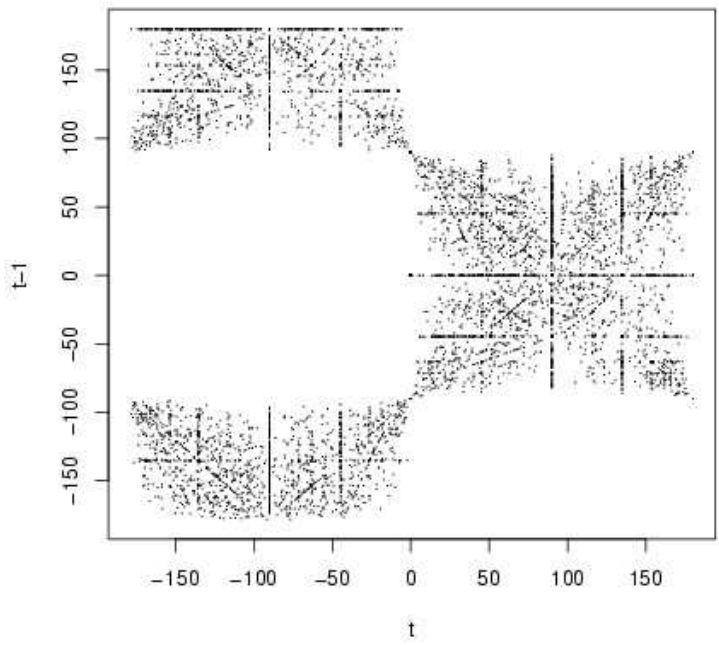}}\\
    \subfigure[Lag 2.]{\includegraphics*[width=4in]{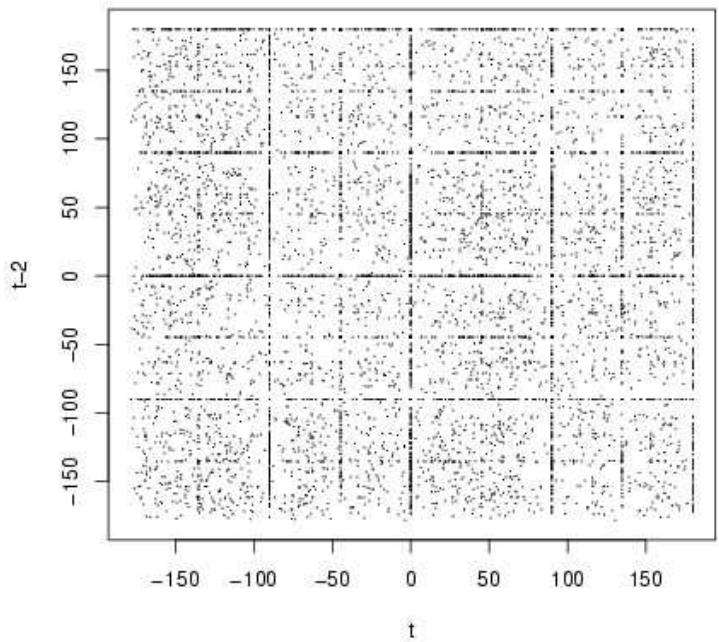}}
  \caption{Patterns observed in the arc phase portraits.}\label{fig:fig2}
\end{figure}

If we plot the arc values $\Phi$ as a time series we also obtain a
pattern that confirms this clustering of points (see Fig.
\ref{fig:fig1}c). Plotting a sorted version of the sequence in
Fig. \ref{fig:fig1}(c), provides us with the pattern observed in
\ref{fig:fig1}(d), where the plateaus indicate more clearly where
the clustering occurs. The intensity of the compass rose patterns
may differ between stocks, however the clustering along the main
directions as discussed above, usually prevails. Searching for
some type of temporal dependency, we looked into the phase
portraits of the arcs. In Fig. \ref{fig:fig2}(a) we have the first
lag phase portrait and in Fig. \ref{fig:fig2}(b) the second lag
one. We can see very curious patterns arising in both diagrams.
There are grids (Fig. \ref{fig:fig2}b) and rays (Fig.
\ref{fig:fig2}a) which correspond to the main rays of the compass
rose as seen in Fig. \ref{fig:fig1}(c) and (d). These patterns
need more analysis.

\medskip
\begin{center}
[ Insert Fig. \ref{fig:fig2} about here. ] \end{center}
\medskip

In order to provide evidence of some kind of dependency in the
above patterns we calculated the BDS test (see \citealp{Brock87}
and \citealp{BrockBaek91}) for the TESCO sequence of arc values
$\Phi$. In table \ref{tab:tab1} we see clearly that for a range of
dimensions $d_E$ (2 to 10), the BDS test reports non IID dynamics
for neighborhood area ranges between 0.5 and 1.5 times the
standard deviation of arcs. When the size becomes twice the
standard deviation (which is regarded as a large radius), we can
accept independence only up to dimension 3. This is an initial
indication that the dynamics as observed from the sequence of
arcs, may contain some dependency that could be used for
forecasting purposes.

\medskip
\begin{center}[ Insert table \ref{tab:tab1} about here. ]\end{center}
\medskip

To back up this result, we calculated the BDS test within a
Surrogate Data Analysis framework (SDA: \citealp{Theiler92:surr}).
SDA is a permutation test framework, similar to bootstrapping, and
is used to test specific nulls that exclude certain dynamics.
Every null comes with its own tailor made simulation procedures
for creating surrogate data sets from the original sequences
(refer to  \citealp{Kaplan95}, \citealp{Kantz},
\citealp{Schreiber96a,Schreiber00} and \citealp{Kugium01} for more
details). In nutshell, via SDA one searches for large
discrepancies between the statistic values on the original and
surrogate sequences. When this occurs,  one can safely reject the
hypothesis that the original sequence stems from a process that is
in compliance with the null. SDA is appropriate here as it allows
us to bypass the limitations of chaotic invariance measures (such
as Lyapunov exponents and dimension statistics) due to small data
sets and noisy information.  Although the BDS is a test for
independence and not nonlinear determinism and complexity, under
the SDA framework it can be used to test for absence of stochastic
randomness via the exclusion of the null hypothesis.

For our case we used the null of the $\Phi$ sequences being
\emph{a monotonic nonlinear transformation of linearly filtered
noise}, which is also regarded as the ``most interesting". This
 implies strongly, the absence of stochastic (random)
dynamics. We analyzed data from 53 FTSE 100 index stocks, spanning
the period 01/01/1970 to 5/30/2003 (a maximum of 8717
observations).  In tables \ref{tab:tab2} and \ref{tab:tab3} we
present the results for 5\% and 2.5\% levels of statistical
significance, for neighborhood sizes $\epsilon$ ranging from 0.5
to 2 times the standard deviation of $\Phi$. We clearly see that
for up to 1.5 times the standard deviation, the null can be safely
rejected (observe the large biases). However, for $\epsilon$
equalling twice the standard deviation, we can not always reject
the null (but this is only for a few cases). Our SDA results  do
not reject the case of forecastability on the basis of arc values.
They also suggest the presence of complex dynamics and the
possibility of some level of determinism.

\medskip
\begin{center}
[ Insert table \ref{tab:tab2} about here. ]\end{center}
\medskip

\medskip
\begin{center}[ Insert table \ref{tab:tab3} about here. ]\end{center}
\medskip

\section{Conclusions}

By adopting an approach similar to \cite{Koppl:01} we showed that
the compass rose \citep{Crack96} can provide useful information
for understanding further the dynamics of stock return sequences.
Moreover, we do not reject the case of these dynamics being
complex and forecastable. We also do not exclude the possibility
of nonlinear determinism. However more research is needed on this
controversial area (see \citealp{Lo:88}, \citealp{MM92},  and
\citealp{Hsieh:91}). An interesting area for future research would
be to experiment with differently simulated processes (such as
purely chaotic, stochastic, (G)ARCH and mixtures of these) and
examine similarities or dissimilarities with the results for the
framework we follow in this letter. It would also be interesting
to see how inhomogeneous sampling of such processes could alter
the results we report here. That could provide useful information
on how our view of the stock return dynamics changes as we move
from high-frequency data to lower frequencies, which is also an
issue that has been concerning the general compass rose
literature.

\ack{The author thanks Enrico Capobianco and the participants of
the Microstructure workshop (organized by CentER, University of
Tilburg, April 2004) for their useful comments and suggestions on
an earlier stage of this work. We also wish to acknowledge the
valuable help and support of Duncan Rand and the University of
Durham ``High Performance Computing Service". The author retains
the sole responsibility of any errors.}

\newpage


\clearpage \newpage \pagestyle{empty} \oddsidemargin  0.0in
    \evensidemargin 0.0in

\newpage

\begin{table}[ht]
  \centering
  \caption{BDS test statistic results and p-values for the TESCO arc values $\Phi$. Embedding dimensions $d_E$ 2 to 10.
   Neighborhood size $\epsilon$ in terms of standard deviation $s$ of $\Phi$.}\label{tab:tab1}
\begin{tabular}{rrrrr}
\hline  \multicolumn{5}{c}{BDS Statistic}  \\
\hline  $\epsilon$: & $0.5 \times s$ & $ 1.0 \times s$ & $ 1.5
\times s$ & $ 2.0  \times s$ \\ \hline \hline
 $d_E$ & 0.897 & 1.794 & 2.691 & 3.588 \\
\hline
2 & 745.31 & 241.38 & 65.39 & 3.45 \\
3 & 846.67 & 223.27 & 56.12 & 2.84 \\
4 & 1010.50 & 220.27 & 49.64 & 1.68 \\
5 & 1243.98 & 222.22 & 44.97 & 0.95 \\
6 & 1577.48 & 229.07 & 41.54 & 0.39 \\
7 & 2053.94 & 239.21 & 38.59 & $-$0.18 \\
8 & 2743.93 & 252.22 & 36.10 & $-$0.72 \\
9 & 3741.75 & 269.69 & 34.16 & $-$1.08 \\
10 & 5201.71 & 291.13 & 32.85 & $-$1.16 \\
\hline
\hline  \multicolumn{5}{c}{BDS Statistic p-values}  \\
\hline
2 & 0.00 & 0.00 & 0.00 & 0.00 \\
3 & 0.00 & 0.00 & 0.00 & 0.00 \\
4 & 0.00 & 0.00 & 0.00 & 0.09 \\
5 & 0.00 & 0.00 & 0.00 & 0.34 \\
6 & 0.00 & 0.00 & 0.00 & 0.69 \\
7 & 0.00 & 0.00 & 0.00 & 0.86 \\
8 & 0.00 & 0.00 & 0.00 & 0.47 \\
9 & 0.00 & 0.00 & 0.00 & 0.28 \\
10 & 0.00 & 0.00 & 0.00 & 0.25 \\
\hline \hline

\end{tabular}

\end{table}

\newpage

\begin{table}[ht]
  \begin{tiny}
  \centering
\caption{Surrogate Data Analysis results on arc values $\Phi$ for
53 companies in the FTSE100. Discriminating statistic: BDS test
(embedding dimension 2). Neighborhood size $\epsilon_1 = 0.5
\times s$, $\epsilon_2 = 1.0 \times s$, $\epsilon_3 = 1.5 \times
s$ and $\epsilon_4 = 2.0 \times s$, where $s =$ standard deviation
of $\Phi$ sequence. Biases and standard errors reported for
significance level $\alpha = 5\%$.}\label{tab:tab2}
  \begin{tabular}{lrrrrrrrrrrrr}
\hline &  \multicolumn{4}{c}{Statistic (BDS)} &
\multicolumn{4}{c}{Bias} & \multicolumn{4}{c}{Standard Error} \\
\hline
Neighborhood size $\epsilon$:  & $\epsilon_1$ & $\epsilon_2$ & $\epsilon_3$ & $\epsilon_4$  & $\epsilon_1$ & $\epsilon_2$ & $\epsilon_3$ & $\epsilon_4$  & $\epsilon_1$ & $\epsilon_2$ & $\epsilon_3$ & $\epsilon_4$ \\
\hline \hline
FTSE ALL SHARE - PRICE INDEX    &   1249.61 &   339.45  &   48.02   &    $-$21.75   &    $-$1246.52     &    $-$338.09  &    $-$47.34   &   22.31   &   2.53    &   1.54    &   1.19    &    1.16 \\
FTSE 100 - PRICE INDEX  &   152.69  &   97.72   &   35.55   &   20.52   &    $-$152.91  &    $-$97.70   &    $-$35.56   &    $-$20.58   &   0.87    &   1.00    &   0.94    &    1.21 \\
ALLIED DOMECQ   &   823.28  &   259.24  &   68.38   &   1.29    &    $-$821.57  &    $-$258.47  &    $-$67.80   &    $-$0.93    &   1.81    &   1.00    &   0.99    &    1.06 \\
AMVESCAP    &   159.36  &   130.18  &   60.63   &   28.46   &    $-$158.57  &    $-$129.63  &    $-$59.80   &    $-$27.64   &   1.20    &   1.24    &   1.26    &    1.25 \\
ASSD.BRIT.FOODS &   290.34  &   163.95  &   62.85   &   21.22   &    $-$289.08  &    $-$162.77  &    $-$61.89   &    $-$20.08   &   1.37    &   1.13    &   1.32    &    1.26 \\
AVIVA   &   928.47  &   270.09  &   66.89   &   0.16    &    $-$927.36  &    $-$269.49  &    $-$66.61   &   0.11    &   1.38    &   0.96    &   1.17    &    1.13 \\
BARCLAYS    &   1076.24 &   281.21  &   62.10   &    $-$5.50    &    $-$1076.56     &    $-$281.26  &    $-$62.33   &   5.20    &   1.21    &   1.16    &   1.09    &    1.13 \\
BOC GROUP   &   829.83  &   257.29  &   67.76   &   3.58    &    $-$828.10  &    $-$256.45  &    $-$67.26   &    $-$3.16    &   1.70    &   1.05    &   0.91    &    0.92 \\
BOOTS GROUP &   1076.14 &   283.31  &   64.05   &    $-$5.67    &    $-$1076.77     &    $-$283.69  &    $-$64.38   &   5.31    &   1.01    &   0.92    &   0.79    &    0.81 \\
BP  &   1115.66 &   295.08  &   63.04   &    $-$7.99    &    $-$1114.58     &    $-$294.74  &    $-$62.90   &   8.04    &   1.32    &   1.04    &   1.24    &    1.26 \\
BRIT.AMERICAN TOBACCO   &   961.19  &   273.01  &   66.07   &    $-$3.63    &    $-$959.84  &    $-$272.50  &    $-$65.74   &   3.80    &   1.23    &   0.82    &   0.75    &    0.76 \\
BRITISH LAND    &   568.83  &   223.68  &   66.58   &   4.10    &    $-$567.36  &    $-$223.00  &    $-$66.02   &    $-$3.65    &   1.02    &   0.94    &   0.72    &    0.74 \\
BUNZL   &   192.12  &   138.33  &   59.76   &   23.61   &    $-$191.62  &    $-$137.83  &    $-$59.07   &    $-$23.05   &   1.01    &   0.97    &   1.18    &    1.10 \\
CADBURY SCHWEPPES   &   653.11  &   233.87  &   70.17   &   6.34    &    $-$651.55  &    $-$232.74  &    $-$69.52   &    $-$5.83    &   1.25    &   0.79    &   0.70    &    0.68 \\
DAILY MAIL 'A'  &   89.23   &   96.50   &   47.69   &   35.62   &    $-$88.25   &    $-$95.23   &    $-$46.43   &    $-$33.71   &   0.78    &   0.95    &   0.85    &    0.77 \\
DIAGEO  &   619.63  &   225.74  &   67.63   &   5.33    &    $-$618.52  &    $-$224.98  &    $-$67.00   &    $-$4.83    &   1.01    &   0.88    &   0.84    &    0.71 \\
DIXONS GP.  &   341.08  &   172.89  &   60.51   &   13.26   &    $-$340.48  &    $-$172.19  &    $-$60.22   &    $-$12.95   &   1.19    &   1.12    &   1.11    &    1.22 \\
EMAP    &   72.70   &   69.19   &   35.03   &   20.02   &    $-$72.73   &    $-$68.86   &    $-$35.12   &    $-$20.23   &   1.26    &   1.27    &   1.26    &    1.07 \\
EXEL    &   433.53  &   203.06  &   69.26   &   9.61    &    $-$433.38  &    $-$203.00  &    $-$69.29   &    $-$9.73    &   1.01    &   0.82    &   0.90    &    0.87 \\
FOREIGN \& COLONIAL &   711.36  &   237.89  &   63.39   &   6.31    &    $-$709.53  &    $-$236.99  &    $-$62.93   &    $-$5.96    &   1.00    &   1.04    &   1.19    &    1.07 \\
GKN &   979.42  &   279.06  &   65.76   &    $-$2.93    &    $-$978.52  &    $-$278.57  &    $-$65.37   &   3.30    &   1.39    &   0.88    &   0.81    &    0.80 \\
GLAXOSMITHKLINE &   1069.02 &   282.23  &   62.10   &    $-$6.11    &    $-$1068.62     &    $-$282.08  &    $-$61.98   &   6.14    &   1.79    &   1.12    &   1.04    &    1.02 \\
GRANADA &   459.67  &   200.60  &   65.15   &   9.49    &    $-$459.69  &    $-$200.59  &    $-$65.08   &    $-$9.41    &   1.27    &   1.19    &   1.11    &    1.16 \\
GUS &   128.72  &   111.26  &   50.83   &   35.48   &    $-$128.49  &    $-$110.76  &    $-$50.36   &    $-$34.92   &   0.89    &   0.92    &   0.93    &    0.83 \\
HANSON  &   804.72  &   251.67  &   61.16   &    $-$2.19    &    $-$804.64  &    $-$251.61  &    $-$61.11   &   2.20    &   1.06    &   1.09    &   0.90    &    0.98 \\
HILTON GROUP    &   777.65  &   248.02  &   70.54   &   4.30    &    $-$775.18  &    $-$246.83  &    $-$69.75   &    $-$3.75    &   1.92    &   1.20    &   1.21    &    1.18 \\
IMP.CHM.INDS.   &   1007.77 &   283.71  &   65.69   &    $-$5.80    &    $-$1007.17     &    $-$283.64  &    $-$65.66   &   5.77    &   1.64    &   0.92    &   1.22    &    1.21 \\
JOHNSON MATTHEY &   378.71  &   178.51  &   63.97   &   14.39   &    $-$378.47  &    $-$178.21  &    $-$63.81   &    $-$14.15   &   0.91    &   0.99    &   0.89    &    0.89 \\
LAND SECURITIES &   1089.61 &   282.28  &   62.79   &    $-$6.22    &    $-$1088.38     &    $-$281.71  &    $-$62.41   &   6.55    &   1.83    &   1.30    &   1.22    &    1.20 \\
LEGAL \& GENERAL    &   970.82  &   269.13  &   65.65   &   0.04    &    $-$969.14  &    $-$268.62  &    $-$65.34   &   0.22    &   1.09    &   1.10    &   0.88    &    0.92 \\
MARKS \& SPENCER GROUP  &   912.70  &   272.35  &   64.73   &    $-$2.03    &    $-$909.93  &    $-$270.77  &    $-$63.62   &   3.04    &   1.67    &   1.36    &   1.32    &    1.26 \\
MORRISON (WM) SPMKTS.   &   90.75   &   90.64   &   45.66   &   44.19   &    $-$89.83   &    $-$89.01   &    $-$44.06   &    $-$41.71   &   0.93    &   0.97    &   0.76    &    1.07 \\
NEXT    &   253.11  &   151.42  &   55.97   &   18.84   &    $-$252.81  &    $-$150.97  &    $-$55.62   &    $-$18.50   &   1.02    &   1.16    &   0.94    &    1.13 \\
PEARSON &   444.27  &   196.80  &   64.87   &   8.11    &    $-$444.02  &    $-$196.49  &    $-$64.80   &    $-$8.04    &   1.08    &   0.84    &   0.94    &    0.84 \\
PROVIDENT FINL. &   266.69  &   155.41  &   58.69   &   16.83   &    $-$266.57  &    $-$155.36  &    $-$58.50   &    $-$16.76   &   0.77    &   0.98    &   1.11    &    0.77 \\
PRUDENTIAL  &   910.07  &   267.31  &   65.78   &   0.13    &    $-$909.09  &    $-$266.63  &    $-$65.29   &   0.37    &   1.08    &   0.98    &   0.79    &    0.79 \\
RECKITT BENCKISER   &   836.20  &   258.41  &   67.44   &   2.10    &    $-$836.20  &    $-$258.47  &    $-$67.68   &    $-$2.30    &   1.33    &   1.29    &   1.19    &    1.24 \\
REED ELSEVIER   &   898.32  &   261.80  &   63.61   &    $-$1.99    &    $-$898.15  &    $-$261.72  &    $-$63.60   &   1.93    &   1.37    &   1.14    &   1.04    &    1.01 \\
RENTOKIL INITIAL    &   197.81  &   131.19  &   56.03   &   28.14   &    $-$196.12  &    $-$129.70  &    $-$54.49   &    $-$26.38   &   0.97    &   0.94    &   0.96    &    0.82 \\
REXAM   &   697.01  &   237.79  &   72.49   &   6.51    &    $-$695.08  &    $-$236.83  &    $-$71.66   &    $-$5.71    &   1.38    &   0.90    &   0.90    &    0.88 \\
RIO TINTO   &   849.08  &   260.87  &   66.48   &    $-$3.08    &    $-$848.51  &    $-$260.71  &    $-$66.40   &   3.19    &   1.59    &   1.37    &   1.33    &    1.31 \\
ROYAL BANK OF SCOTLAND  &   433.58  &   198.72  &   65.90   &   9.33    &    $-$432.41  &    $-$198.07  &    $-$65.28   &    $-$8.83    &   0.88    &   1.06    &   0.93    &    0.86 \\
SAINSBURY (J)   &   698.76  &   227.30  &   59.38   &   0.21    &    $-$698.30  &    $-$227.22  &    $-$59.18   &    $-$0.21    &   1.15    &   1.29    &   1.29    &    1.21 \\
SCHRODERS   &   72.28   &   84.11   &   40.53   &   41.80   &    $-$71.93   &    $-$83.50   &    $-$39.86   &    $-$40.50   &   1.14    &   1.17    &   1.16    &    1.20 \\
SCOT. \& NEWCASTLE  &   750.36  &   245.66  &   67.65   &   0.87    &    $-$749.30  &    $-$244.78  &    $-$67.08   &    $-$0.40    &   1.09    &   0.72    &   0.94    &    0.95 \\
SHELL TRANSPORT \& TRDG.    &   1199.79 &   294.59  &   65.20   &    $-$6.32    &    $-$1198.76     &    $-$293.78  &    $-$64.65   &   6.72    &   1.46    &   1.07    &   1.04    &    1.04 \\
SMITH \& NEPHEW &   500.14  &   208.90  &   71.11   &   14.25   &    $-$497.28  &    $-$206.89  &    $-$69.41   &    $-$12.80   &   1.64    &   1.67    &   1.60    &    1.60 \\
SMITHS GROUP    &   587.34  &   222.17  &   69.99   &   10.48   &    $-$586.92  &    $-$222.11  &    $-$70.07   &    $-$10.66   &   1.36    &   0.92    &   1.27    &    1.29 \\
STD.CHARTERED   &   623.16  &   223.03  &   67.43   &   5.56    &    $-$622.41  &    $-$222.38  &    $-$66.93   &    $-$5.16    &   1.24    &   1.05    &   0.84    &    0.93 \\
TESCO   &   745.31  &   241.38  &   65.24   &   3.24    &    $-$744.56  &    $-$240.87  &    $-$64.96   &    $-$2.97    &   1.55    &   1.15    &   1.15    &    1.14 \\
TOMKINS &   77.66   &   73.89   &   37.40   &   25.49   &    $-$77.65   &    $-$73.70   &    $-$37.12   &    $-$25.38   &   1.17    &   1.29    &   0.94    &    0.93 \\
UNILEVER (UK)   &   1017.26 &   278.21  &   59.93   &    $-$7.77    &    $-$1017.15     &    $-$278.20  &    $-$59.90   &   7.76    &   1.39    &   1.04    &   1.09    &    0.98 \\
WHITBREAD   &   851.02  &   262.03  &   66.73   &   3.23    &    $-$849.13  &    $-$260.92  &    $-$65.95   &    $-$2.46    &   1.79    &   1.20    &   1.11    &    1.11 \\
WOLSELEY    &   172.73  &   129.30  &   52.87   &   27.14   &    $-$172.66  &    $-$129.15  &    $-$52.69   &    $-$27.06   &   0.82    &   0.95    &   1.03    &    0.89 \\
WPP GROUP   &   71.43   &   60.94   &   38.30   &   21.64   &    $-$71.01   &    $-$60.48   &    $-$37.74   &    $-$21.25   &   1.11    &   0.81    &   0.82    &    0.82 \\
\hline \hline
  \end{tabular}
  \end{tiny}
\end{table}

\clearpage

\begin{table}[ht]
  \begin{tiny}
  \centering
\caption{Surrogate Data Analysis results on arc values $\Phi$ for
53 companies in the FTSE100. Discriminating statistic: BDS test
(embedding dimension 2). Neighborhood size $\epsilon_1 = 0.5
\times s$, $\epsilon_2 = 1.0 \times s$, $\epsilon_3 = 1.5 \times
s$ and $\epsilon_4 = 2.0 \times s$, where $s =$ standard deviation
of $\Phi$ sequence. Biases and standard errors reported for
significance level $\alpha = 2.5\%$.}\label{tab:tab3}
  \begin{tabular}{lrrrrrrrrrrrr}
\hline &  \multicolumn{4}{c}{Statistic (BDS)} &
\multicolumn{4}{c}{Bias} & \multicolumn{4}{c}{Standard Error} \\
\hline
Neighborhood size $\epsilon$:  & $\epsilon_1$ & $\epsilon_2$ & $\epsilon_3$ & $\epsilon_4$  & $\epsilon_1$ & $\epsilon_2$ & $\epsilon_3$ & $\epsilon_4$  & $\epsilon_1$ & $\epsilon_2$ & $\epsilon_3$ & $\epsilon_4$ \\
\hline \hline
FTSE ALL SHARE - PRICE INDEX    &   1249.61 &   339.45  &   48.02   &    $-$21.75   &    $-$1245.98     &    $-$338.20  &    $-$47.50   &   22.15   &   1.97    &   0.98    &   0.98    &    1.01 \\
FTSE 100 - PRICE INDEX  &   152.69  &   97.72   &   35.55   &   20.52   &    $-$152.69  &    $-$97.82   &    $-$35.55   &    $-$20.54   &   0.98    &   0.93    &   0.86    &    1.12 \\
ALLIED DOMECQ   &   823.28  &   259.24  &   68.38   &   1.29    &    $-$821.92  &    $-$258.54  &    $-$68.02   &    $-$1.07    &   1.35    &   0.96    &   0.97    &    0.92 \\
AMVESCAP    &   159.36  &   130.18  &   60.63   &   28.46   &    $-$158.41  &    $-$129.22  &    $-$59.67   &    $-$27.18   &   0.90    &   0.83    &   0.85    &    0.89 \\
ASSD.BRIT.FOODS &   290.34  &   163.95  &   62.85   &   21.22   &    $-$288.64  &    $-$162.40  &    $-$61.51   &    $-$19.84   &   1.35    &   1.33    &   1.37    &    1.30 \\
AVIVA   &   928.47  &   270.09  &   66.89   &   0.16    &    $-$926.88  &    $-$269.44  &    $-$66.56   &   0.05    &   1.36    &   1.18    &   1.23    &    1.17 \\
BARCLAYS    &   1076.24 &   281.21  &   62.10   &    $-$5.50    &    $-$1075.93     &    $-$281.31  &    $-$62.19   &   5.38    &   1.36    &   1.14    &   1.16    &    1.11 \\
BOC GROUP   &   829.83  &   257.29  &   67.76   &   3.58    &    $-$828.19  &    $-$256.50  &    $-$67.12   &    $-$3.08    &   1.62    &   1.03    &   0.89    &    0.84 \\
BOOTS GROUP &   1076.14 &   283.31  &   64.05   &    $-$5.67    &    $-$1075.91     &    $-$283.18  &    $-$63.98   &   5.69    &   1.26    &   1.01    &   0.95    &    0.92 \\
BP  &   1115.66 &   295.08  &   63.04   &    $-$7.99    &    $-$1114.62     &    $-$294.54  &    $-$62.78   &   8.18    &   1.70    &   1.34    &   1.19    &    1.20 \\
BRIT.AMERICAN TOBACCO   &   961.19  &   273.01  &   66.07   &    $-$3.63    &    $-$959.60  &    $-$272.40  &    $-$65.67   &   3.99    &   1.37    &   1.08    &   1.00    &    0.94 \\
BRITISH LAND    &   568.83  &   223.68  &   66.58   &   4.10    &    $-$567.30  &    $-$222.67  &    $-$65.59   &    $-$3.19    &   1.17    &   0.95    &   1.03    &    1.00 \\
BUNZL   &   192.12  &   138.33  &   59.76   &   23.61   &    $-$191.61  &    $-$137.79  &    $-$59.05   &    $-$22.97   &   0.99    &   0.98    &   0.90    &    0.84 \\
CADBURY SCHWEPPES   &   653.11  &   233.87  &   70.17   &   6.34    &    $-$651.52  &    $-$232.88  &    $-$69.47   &    $-$5.61    &   1.35    &   1.04    &   1.02    &    1.00 \\
DAILY MAIL 'A'  &   89.23   &   96.50   &   47.69   &   35.62   &    $-$88.45   &    $-$95.38   &    $-$46.60   &    $-$33.87   &   1.07    &   0.99    &   1.01    &    1.14 \\
DIAGEO  &   619.63  &   225.74  &   67.63   &   5.33    &    $-$618.36  &    $-$225.24  &    $-$67.24   &    $-$5.02    &   1.26    &   1.15    &   1.19    &    1.22 \\
DIXONS GP.  &   341.08  &   172.89  &   60.51   &   13.26   &    $-$340.07  &    $-$172.18  &    $-$59.87   &    $-$12.58   &   0.97    &   0.99    &   0.88    &    0.99 \\
EMAP    &   72.70   &   69.19   &   35.03   &   20.02   &    $-$72.78   &    $-$68.80   &    $-$34.77   &    $-$20.02   &   1.04    &   1.08    &   0.87    &    1.04 \\
EXEL    &   433.53  &   203.06  &   69.26   &   9.61    &    $-$433.28  &    $-$202.75  &    $-$69.08   &    $-$9.38    &   1.04    &   1.04    &   0.99    &    0.97 \\
FOREIGN \& COLONIAL &   711.36  &   237.89  &   63.39   &   6.31    &    $-$709.87  &    $-$236.98  &    $-$62.74   &    $-$5.78    &   1.23    &   1.21    &   1.03    &    1.08 \\
GKN &   979.42  &   279.06  &   65.76   &    $-$2.93    &    $-$978.64  &    $-$278.57  &    $-$65.46   &   3.20    &   1.33    &   1.10    &   1.04    &    1.08 \\
GLAXOSMITHKLINE &   1069.02 &   282.23  &   62.10   &    $-$6.11    &    $-$1068.51     &    $-$281.94  &    $-$61.86   &   6.30    &   1.29    &   1.03    &   1.03    &    1.04 \\
GRANADA &   459.67  &   200.60  &   65.15   &   9.49    &    $-$459.69  &    $-$200.39  &    $-$65.16   &    $-$9.38    &   1.15    &   1.29    &   1.14    &    1.22 \\
GUS &   128.72  &   111.26  &   50.83   &   35.48   &    $-$128.32  &    $-$110.95  &    $-$50.35   &    $-$34.85   &   1.15    &   1.12    &   1.14    &    1.09 \\
HANSON  &   804.72  &   251.67  &   61.16   &    $-$2.19    &    $-$804.75  &    $-$251.79  &    $-$61.30   &   2.07    &   1.09    &   1.03    &   0.99    &    1.02 \\
HILTON GROUP    &   777.65  &   248.02  &   70.54   &   4.30    &    $-$776.09  &    $-$247.12  &    $-$69.98   &    $-$3.90    &   1.31    &   1.17    &   1.10    &    1.14 \\
IMP.CHM.INDS.   &   1007.77 &   283.71  &   65.69   &    $-$5.80    &    $-$1007.48     &    $-$283.85  &    $-$65.79   &   5.67    &   1.33    &   1.03    &   0.88    &    0.88 \\
JOHNSON MATTHEY &   378.71  &   178.51  &   63.97   &   14.39   &    $-$378.02  &    $-$177.88  &    $-$63.54   &    $-$13.91   &   1.03    &   1.01    &   0.89    &    0.88 \\
LAND SECURITIES &   1089.61 &   282.28  &   62.79   &    $-$6.22    &    $-$1088.71     &    $-$281.87  &    $-$62.38   &   6.55    &   1.26    &   1.10    &   0.99    &    0.95 \\
LEGAL \& GENERAL    &   970.82  &   269.13  &   65.65   &   0.04    &    $-$969.44  &    $-$268.38  &    $-$65.24   &   0.33    &   1.36    &   1.08    &   0.96    &    0.96 \\
MARKS \& SPENCER GROUP  &   912.70  &   272.35  &   64.73   &    $-$2.03    &    $-$909.91  &    $-$270.94  &    $-$63.82   &   2.77    &   1.33    &   0.95    &   0.78    &    0.85 \\
MORRISON (WM) SPMKTS.   &   90.75   &   90.64   &   45.66   &   44.19   &    $-$90.24   &    $-$89.74   &    $-$44.68   &    $-$42.54   &   0.93    &   0.90    &   0.92    &    0.93 \\
NEXT    &   253.11  &   151.42  &   55.97   &   18.84   &    $-$252.66  &    $-$151.03  &    $-$55.45   &    $-$18.50   &   0.82    &   0.96    &   0.82    &    0.73 \\
PEARSON &   444.27  &   196.80  &   64.87   &   8.11    &    $-$443.83  &    $-$196.62  &    $-$64.64   &    $-$7.87    &   1.09    &   1.02    &   1.00    &    0.94 \\
PROVIDENT FINL. &   266.69  &   155.41  &   58.69   &   16.83   &    $-$266.55  &    $-$155.49  &    $-$58.70   &    $-$16.89   &   1.04    &   1.29    &   1.20    &    1.09 \\
PRUDENTIAL  &   910.07  &   267.31  &   65.78   &   0.13    &    $-$909.18  &    $-$267.05  &    $-$65.59   &    $-$0.00    &   1.22    &   1.01    &   1.06    &    1.11 \\
RECKITT BENCKISER   &   836.20  &   258.41  &   67.44   &   2.10    &    $-$835.56  &    $-$257.93  &    $-$67.04   &    $-$1.72    &   1.20    &   1.12    &   1.00    &    0.98 \\
REED ELSEVIER   &   898.32  &   261.80  &   63.61   &    $-$1.99    &    $-$897.92  &    $-$261.77  &    $-$63.47   &   2.14    &   1.37    &   1.13    &   1.12    &    1.13 \\
RENTOKIL INITIAL    &   197.81  &   131.19  &   56.03   &   28.14   &    $-$196.05  &    $-$129.48  &    $-$54.33   &    $-$26.04   &   0.91    &   1.19    &   1.09    &    1.17 \\
REXAM   &   697.01  &   237.79  &   72.49   &   6.51    &    $-$695.26  &    $-$236.76  &    $-$71.64   &    $-$5.70    &   1.53    &   1.21    &   1.17    &    1.21 \\
RIO TINTO   &   849.08  &   260.87  &   66.48   &    $-$3.08    &    $-$848.18  &    $-$260.53  &    $-$66.23   &   3.30    &   1.03    &   0.98    &   0.85    &    0.86 \\
ROYAL BANK OF SCOTLAND  &   433.58  &   198.72  &   65.90   &   9.33    &    $-$432.32  &    $-$198.01  &    $-$65.34   &    $-$8.82    &   1.15    &   0.96    &   0.81    &    0.83 \\
SAINSBURY (J)   &   698.76  &   227.30  &   59.38   &   0.21    &    $-$698.48  &    $-$227.33  &    $-$59.46   &    $-$0.22    &   0.93    &   0.95    &   0.83    &    0.82 \\
SCHRODERS   &   72.28   &   84.11   &   40.53   &   41.80   &    $-$71.77   &    $-$83.11   &    $-$39.59   &    $-$40.11   &   1.03    &   1.13    &   1.10    &    1.26 \\
SCOT. \& NEWCASTLE  &   750.36  &   245.66  &   67.65   &   0.87    &    $-$749.08  &    $-$245.00  &    $-$67.06   &    $-$0.26    &   1.03    &   0.96    &   0.91    &    0.89 \\
SHELL TRANSPORT \& TRDG.    &   1199.79 &   294.59  &   65.20   &    $-$6.32    &    $-$1198.60     &    $-$294.14  &    $-$65.03   &   6.45    &   1.58    &   1.17    &   1.10    &    1.07 \\
SMITH \& NEPHEW &   500.14  &   208.90  &   71.11   &   14.25   &    $-$497.73  &    $-$207.15  &    $-$69.72   &    $-$12.99   &   1.17    &   1.20    &   1.06    &    1.11 \\
SMITHS GROUP    &   587.34  &   222.17  &   69.99   &   10.48   &    $-$586.43  &    $-$221.51  &    $-$69.40   &    $-$9.90    &   1.28    &   0.97    &   1.12    &    1.14 \\
STD.CHARTERED   &   623.16  &   223.03  &   67.43   &   5.56    &    $-$623.18  &    $-$222.82  &    $-$67.38   &    $-$5.56    &   1.18    &   1.10    &   1.13    &    1.21 \\
TESCO   &   745.31  &   241.38  &   65.24   &   3.24    &    $-$745.08  &    $-$241.37  &    $-$65.23   &    $-$3.33    &   1.12    &   1.16    &   1.10    &    1.08 \\
TOMKINS &   77.66   &   73.89   &   37.40   &   25.49   &    $-$77.72   &    $-$73.55   &    $-$37.02   &    $-$25.33   &   0.87    &   0.98    &   1.10    &    0.90 \\
UNILEVER (UK)   &   1017.26 &   278.21  &   59.93   &    $-$7.77    &    $-$1017.42     &    $-$278.30  &    $-$60.03   &   7.68    &   1.28    &   0.98    &   0.98    &    0.96 \\
WHITBREAD   &   851.02  &   262.03  &   66.73   &   3.23    &    $-$849.28  &    $-$261.35  &    $-$66.25   &    $-$2.81    &   1.44    &   1.10    &   1.12    &    1.16 \\
WOLSELEY    &   172.73  &   129.30  &   52.87   &   27.14   &    $-$172.58  &    $-$129.19  &    $-$52.74   &    $-$26.87   &   1.06    &   1.08    &   1.10    &    0.96 \\
WPP GROUP   &   71.43   &   60.94   &   38.30   &   21.64   &    $-$71.32   &    $-$60.89   &    $-$38.02   &    $-$21.43   &   1.00    &   1.01    &   0.73    &    0.94 \\

\hline \hline
  \end{tabular}
  \end{tiny}
\end{table}

\clearpage

\newpage

\end{document}